\documentstyle[aps,epsfig]{revtex}
\begin{document}
\draft
\title{On possible $\Theta$ vacua states in heavy ion collisions}
\author{\bf A. K. Chaudhuri\cite{byline}}
\address{ Variable Energy Cyclotron Centre\\
1/AF,Bidhan Nagar, Calcutta - 700 064\\}
\noindent
\maketitle

\begin{abstract}
We have simulated the possible $\Theta$ vacua states in heavy ion
collisions.  In  a  quench  like  scenario,  random phases of the
chiral fields  were  evolved  in  a  zero  temperature  potential
incorporating  the  breaking of $U_A(1)$ symmetry. Initial random
phases very quickly settles into oscillation  around  the  values
dictated  by the potential. The simulation indicate that $\Theta$
vacua states that can be populated in heavy ion collisions is 
a  coherent superposition of  a  number  of  modes.
\end{abstract}


QCD  in  the  chiral limit ($m_u=m_d=m_s=0$) possesses a $U_A(1)$
symmetry. Spontaneous breaking of the symmetry require a  neutral
pseudoscalar  Goldstone  boson  with  mass  less  than  $\sqrt{3}
m_\pi$, in addition to pion itself.  However  no  such  Goldstone
boson  is  seen  in  nature.  The  problem  was  resolved  by the
discovery of non-perturbative effcets  that  violates  the  extra
$U_A(1)$  symmetry. 't Hooft showed that because of the instanton
solution of the Yang-Mills  theory,  the  U(1)  symmetry  is  not
really  a  symmetry  of the vacuum \cite{thooft}. Term containing
the so called vacuum angle  ($\Theta$)  which  breaks  P  and  CP
symmetry  can be added to the Lagrangian. QCD requires very small
$\Theta \sim 10^{-9}$,  which  explain  the  apparent  P  and  CP
symmetry   in  strong  interaction.  Dynamical  breaking  of  the
$U_A(1)$ symmetry is also obtained in large N  (color)  limit  of
the  SU(N)  gauge  theory \cite{wi79,wi80,ve80}. In this approach
the dominant fluctuations are not semi-classical but  of  quantum
nature.

Recently  Kharzeev,  Pisarski  and Tytgat \cite{kh98,kh99} argued
that in heavy ion collisions non trivial  $\Theta$  vacua  states
may be created. In the limit of large number of colors, the axial
$U_A(1)$  symmetry  of massless quarks may be restored at the the
deconfining  phase  transition.  As  the  system  rolls  back  to
confining  phase, it may settles into a metastable state with non
trivial $\Theta$.  The  idea  is  similar  to  the  formation  of
Disoriented chiral condensate (DCC) \cite{dcc}. In DCC space-time
region  is  created  where  the  chiral  condensate  points  in a
direction different from that of the ground state. Similarly,  in
$\Theta$ vacua states space-time region with non-trivial $\Theta$
may be created.

After  the  suggestion of Kharzeev et al \cite{kh98,kh99} several
authors have looked into various aspects of non-trivial  $\Theta$
vacua   state   that  may  be  formed  in  heavy  ion  collisions
\cite{ah00,vo00,bu00a,bu00b}. Buckley  et  al  \cite{bu00a,bu00b}
numerically  simulated  the  formation  of $\Theta$ vacua states,
using  the  effective  Lagrangian  of  Halperin  and   Zhitnitsky
\cite{ha98}.  They  assumed quench like scenario, rapid expansion
of the fireball leave  behind  an  effectively  zero  temperature
region  in  the  interior which is isolated from the true vacuum.
Starting from an initial non-equilibrium state, they studied  the
evolution  of  phases  of the chiral field. They saw formation of
non-zero $\Theta$ vacuum within a time scale of  $10^{-23}$  sec.
In  their  formulation they used a dissipative term with friction
constant $\gamma$. $\gamma$=200 MeV  was  chosen,  which  may  be
rather   large  as  the  fields  evolve  essentially  in  a  zero
temperature potential, where dissipation will be small. They also
choose  to  ignore  the  Fluctuation-dissipation  theorem   which
require that dissipation be associated with fluctuations (noise).
They did not include such a term.

In  the  present  paper  we  follow  essentially  the approach of
Buckley et al  \cite{bu00a,bu00b}  to  investigate  the  possible
creation  of  $\Theta$  vacua states in heavy ion collision, with
some important  differences.  We  use  the  effective  Lagrangian
developed  by Witten \cite{wi80} to discuss the $U_A(1)$ anamoly,
and choose to omit the arbitrary dissipative term.  In  a  quench
like  scenario,  chiral phases are evolving in a zero temperature
potential, where dissipative effects are supposed  to  be  small.
Also,  neglecting dissipative term we are avoiding the problem of
using a fluctuation term, non-inclusion of which will violate the
fluctuation-dissipation theorem. We have also choose to  use  the
proper  time  ($\tau$)  and  rapidity (Y) as the most appropriate
coordinate system for heavy ion collisions.

Effective  non-linear sigma model which incorporates the breaking
of $U_A(1)$ symmetry can be written as \cite{wi80},

\begin{equation}
{\cal{L}}=f^2_\pi   (\frac{1}{2}   tr   (\partial_\mu   U^\dagger
\partial_\mu U) + tr(M(U+U^\dagger)) - a (tr \ln U - \Theta)^2)
\label{1}
\end{equation}

\noindent where U is a $3\times 3$ unitary matrix with expansion,
$U=U_0  (1  +  i \sum t^a \pi^a /f_\pi + O (\pi^2))$, $U_0$ being
the vacuum expectation value of U, $t^a$ are  the  generators  of
U(3)  ($Tr  t^a  t^b  =\delta  ^{ab})$ and $\pi^a$ are the nonet of
Goldstone boson fields. M is the  quark  mass  matrix,  which  is
positive,  real  and  diagonal. We denote the diagonal entries as
$\mu_i^2$. They are the Goldstone boson  sqared  masses,  if  the
anamoly  term  $a$  (a/N in ref.\cite{wi80}) were absent. Because
$M$  is  diagonal  $U$  can  be  assumed   to   be   diagonal   ,
$U_{ij}=e^{i\phi_i} \delta_{ij}$.

In terms of $\phi_i$'s, the potential is,

\begin{equation}
V(\phi_i)=f^2_\pi  (-\sum  \mu^2_i  \cos \phi_i +a/2(\sum \phi_i -
\Theta)^2)  \label{2}
\end{equation}

It  may  be noted that as $\sum \phi_i$ arose from $tr \ln U$, it
is  defined  modulo  $2\pi$.  In  the   present   work   we   use
$\mu^2_1=(114MeV)^2$, $\mu^2_2=(161MeV)^2$ ,$\mu^2_3=(687)^2$ and
$a=(492MeV)^2$  \cite{kh98,kh99}. With these parameters, the mass
matrix in (\ref{2}) can be diagonalised to obtain $m_\pi^0  \sim$
139  MeV,  $m_\eta \sim $ 501 MeV and $m_{\eta^\prime} \sim $ 983
MeV, close to their experimental values.

Vacuum  expectation  values  of  the  angles  $\phi_i$'s  can  be
obtained from the minimisation condition,

\begin{equation}
\mu^2_i \sin \phi_i = a (\Theta - \sum \phi_j)
\label{3}
\end{equation}

Solutions of this non-linear coupled equations has been discussed
in  detail  by  Witten  \cite{wi80}. If the equation has only one
solution then physics will be analytic as a function of $\Theta$.
The solution  vary  periodically  in  $\Theta$  with  periodicity
$2\pi$  \cite{wi80}.  Also  this  solution  must be CP conserving
whenever CP is a symmetry of the equation  \cite{wi80}.  However,
it  may  happen  that eq.\ref{3} has more than one solution. Then
the solution are not CP conserving, rather  a  CP  transformation
exchanges them.

Existence  of  metastable  states  ($\Theta$ vacua states) can be
argued as follows: the vacuum expectation values  depend  on  the
ratio  $a/\mu^2_i$, which in the mean field theory decreases with
temperature $a/\mu^2_i \sim (T_d -  T)^{3/2}$,  $T_d$  being  the
decoupling temperature \cite{kh99}. Then as the system rolls back
towards  the  chiral  symmetry breaking, $a/\mu^2_i$ may be small
enough to support metastable states. In  ref.\cite{kh98}  it  was
shown  that  when  $a/\mu^2_1  <  .2467$  there  is  a metastable
solution, which is unstable in  $\pi^0$  direction  unless  $a  <
a_{cr}$, $a_{cr}/mu^2_1\sim .2403$.

Appropriate  coordinates  for heavy ion scattering are the proper
time  ($\tau$)  and  rapidity  (Y).  Assuming  boost  invariance,
equation of motion for the phases $\phi_i$'s can be written as,

\begin{equation}
\frac{\partial^2    \phi_i}{\partial    \tau^2}   +\frac{1}{\tau}
\frac{\partial    \phi_i}{\partial    \tau}     -\frac{\partial^2
\phi_i}{\partial  x^2}  - \frac{\partial^2 \phi_i}{\partial y^2}+
\sum \mu^2_i sin(\phi_i) - a(\sum \phi_i - \Theta)) =0
\label{4}
\end{equation}

It  is  interesting  to  note  that  in this coordinate system, a
dissipative term which decreases with (proper) time is effective.
We  have  solved  the  coupled  partial  differential   equations
assuming  a  quench  like  scenario  after the symmetry restoring
phase transition. The rapid expansion of the  high  energy  shell
leaves  behind  an  effectively zero temperature region, which is
isolated from the true vacuum. The chirally symmetric fields then
essentially evolves in the zero temperature potential.

The  set  of  partial equations (\ref{4}) were solved on a $64^2$
lattice, with lattice spacing of $a=1 fm$ and  time  interval  of
$a/10$  fm.  We  also  use  periodic  boundary condition. Solving
eq.\ref{4} require field configuration at the  initial  time.  We
assume   the  initial  time  as  $\tau_i$=1  fm.  Initial  phases
($\phi_i$ and $\dot{\phi_i}$) were chosen according to  following
prescription,

\begin{eqnarray}
\phi_i = &R_N/(1 + exp((r-r_{cr})/\Gamma) \\
\dot{\phi_i} = & R_N/(1 +exp((r-r_{cr})/\Gamma)
\end{eqnarray}

\noindent  where  $R_N$  is  a  random number within the interval
$[-2\pi/16,2\pi/16]$. We use $r_{cr}$=20 fm and $\Gamma$=.5 fm.

The   phases   were  evolved  according  to  the  eq.\ref{4}  for
arbitrarily long time 50 fm in the  zero  temperature  potential.
The  idea  is  to  see  whether  nontrivial $\Theta$ vacua emerge
during the evolution.  We  have  considered  two  situation,  (i)
$\Theta$=0  and (ii)$\Theta$=$4\pi/16$ within the radius $r_{cr}$
and zero beyond that radius. For $\Theta$=0 emergence of non-zero
$\phi$'s will be signature of $\Theta$ vacua.

In  fig.  1,  the  evolution of the spatially averaged phases are
shown,

\begin{equation}
\phi_i =\frac{\int \phi_i(x,y) dx dy}{\int dx dy}, i=u,d,s
\end{equation}

For $\Theta$ =0, the potential is minimised for $\phi_u= \phi_d =
\phi_s=0$.   The  averaged  fields  also  fluctuate  around  zero
througout the  evolution.  This  shows  that  chirally  symmetric
fields  donot  evolve  into  a  non-trivial  $\Theta$  vacua. The
situation is different for finite $\Theta$. For  $\Theta=4\pi/16$
the potential is minimised for $\phi_u$=0.502, $\phi_d$=0.243 and
$\phi_s$=0.013.  Chirally  symmetric  fields very quickly reaches
these values and oscillate around it. The  oscillation  continues
for  long,  with  little  damping.  As  will  be shown later, the
continues oscillation of the spatially averaged  phases  indicate
that  a  large  number  of modes contribute to the $\Theta$ vacua
state.

We define a correlation function,

\begin{equation}
C(r)=  \frac{\sum_{i,j}  \roarrow{  \phi_i}  .  \roarrow{\phi_j}}
{\sum_{i,j}
|\roarrow{\phi_i}| |\roarrow{\phi_j}|}
\end{equation}

\noindent  such  that the distance between the lattice points $i$
and  $j$  is  $r$.  $C(r)$  specifies  how  the  three  component
$\roarrow{\phi}=(\phi_u,phi_d,\phi_s)$  at two lattice points are
correlated.  In  fig.2,  we  have  shown  the  evolution  of  the
correlation   function   for  $\Theta$=0  and  $\Theta$=0.785  at
different  times,  $\tau$=1,6,11,16  and  21  fm.  Initially   at
$\tau$=1  fm,  for  both the cases, correlation length is about 1
fm, the lattice spacing indicating the fact that  initial  fields
were   random,   with  no  correlation.  The  correlation  length
increases at later times. But for  $\Theta$=0,  the  increase  is
marginal.   However   for  finite  $\Theta$,  correlation  length
increases  very  rapidly,  and  assumes   quite   large   values.
Physically, for finite $\Theta$ all the $\roarrow{\phi}$'s try to
align   themselves  in  some  direction  (i.e.  in  the  $\Theta$
direction) thereby giving very large correlation length even when
they are separated by  a  large  distance.  This  figure  clearly
demonstrated  the  possibility  of  parity odd bubbles formation.
Initially random phases evolve such that they points in the  same
direction, forming a large parity odd bubble.

To  compare  our results with that of Buckley et al \cite{bu00a},
we have also studied the evolution of the different modes of  the
phases  $\phi_i$.  At  each  alternate time step, we apply a fast
fourier transform to the spatial data.  The  fourier  transformed
data  are  then  integrated  over  the  angles to obtain momentum
distribution,

\begin{equation}
\phi_i(k)=\frac{1}{2\pi}  \int^{2\pi}_0 \phi_i(k,\theta) d\theta,
i=u,d,s
\end{equation}

Each  mode  was  averaged over some narrow bin. In fig.3, we have
shown the evolution of $\phi_u$ for $\Theta$=0 (upper panel)  and
$\Theta$=0.785  (lower  panel).  For  both  the  cases, the modes
oscillate throughout the evolution. Initially  amplitude  of  the
oscillation  decreases, but at later time it remains more or less
same. This is  indicative  of  the  fact  that  dissipative  term
decreases   with  time.  We  also  find  that  higher  modes  are
suppressed compared to zero mode, but the suppression is  not  as
large as obtained by Buckley et al \cite{bu00a,bu00b}. Even after
a  large  interval  of time, higher modes donot become negligibly
small. For $\Theta$=0, throughout the evolution, the zero mode as
well as higher modes, oscillate about zero. But for finite  value
of  $\Theta=4\pi/16$  the  modes  oscillate  about  some definite
nonzero value.  For  $\phi_d$  and  $\phi_s$,  we  have  obtained
qualitatively  similar results. Indeed the evolution of different
modes are in accordance with  fig.1,  where  we  have  shown  the
evolution  of  spatial averaged phases. Spatially averaged phases
were found to oscillate about some fixed value  dictated  by  the
minimum   of   the  potential.  Present  simulation  results  are
qualitatively different from the simulation results of Buckley et
al \cite{bu00a,bu00b}. They found that for finite  $\Theta$,  the
modes  quickly reaches the value dictated by the $\Theta$ vacuum.
The difference is essentially due to the strong dissipative  term
in the equation of motion of Buckley et al \cite{bu00a}, which we
have  omitted.  We  have  checked that if we include an arbitrary
dissipative term, zero mode  as  well  as  higher  modes  quickly
evolve  into  a  constant  value,  dictated by the minimum of the
potential.  Also,  in   the   simulation   of   Buckley   et   al
\cite{bu00a,bu00b}, the $\Theta$ vacua is essentially consists of
zero  mode,  while we find that in absence of dissipation, higher
modes also contribute substantially. The $\Theta$ vacua is thus a
coherent a superposition of a number of modes.

Present   simulation   indicate   that  in  heavy  ion  collision
non-trivial $\Theta$-vacua state that is a coherent superposition
of a number modes can be formed. What will be  the  signature  of
such  a  state.  As such detecting $\Theta$ states are difficult.
Kharzeev and Pisarski \cite{kh99} estimated the P-odd observables
are on the order of $10^{-3}$, a small effect. Also as  discussed
by  Voloshin  \cite{vo00} the so called signal of $\Theta$ states
may be faked by "conventional" effect such  as  anisotropic  flow
etc.  Whatever be the signal of the $\Theta$ vacua states, with a
large number of modes contributing,  signal  will  be  broadened,
effectively diluting the detection probability.

To  summarise,  we  have  simulated  the  non-linear  sigma model
incorporating the dynamical  $U_A(1)$  breaking.  Assuming  boost
invariance equation of motion of the chiral phases were solved on
a $64^2$ lattice  with  lattice spacing a=1 fm. It was shown that
phases of chiral fields with finite $\Theta$ oscillate around the
values dictated by the  minimum  of  the  potential.  Correlation
studies  shows  that  initially  uncorrelated phases very quickly
develops large correlation  length,  indicating  formation  of  a
'Parity Odd' bubbles. Fourier analysis of the modes indicate that
the  $\Theta$ vacua states are coherent superposition of a number
of modes, which continue to oscillate with time.

\begin{figure}
\centerline{\psfig{figure=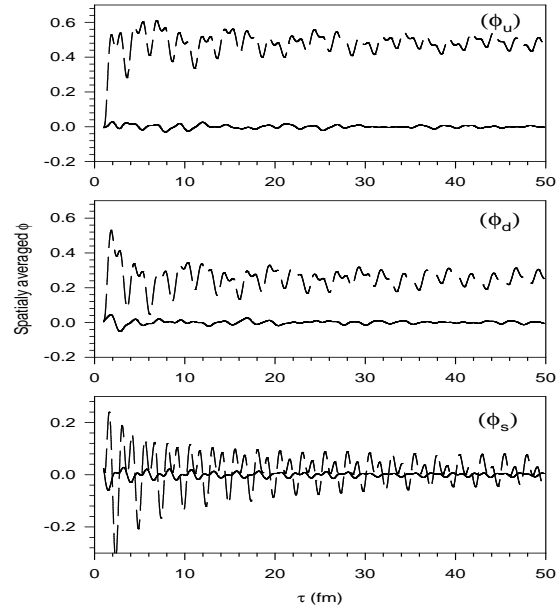,height=10cm,width=10cm}}
\caption{Evolution  of  spatially  averaged $\phi_i$, i=u,d and s
for $\Theta$=0 (solid line) and $\Theta$=0.785 (dashed line).}
\end{figure}
\begin{figure}
\centerline{\psfig{figure=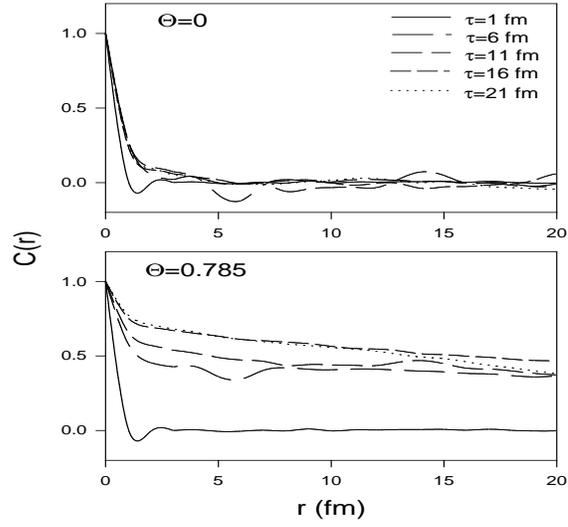,height=10cm,width=10cm}}
\caption{Correlation  function  at different times for $\Theta$=0
(upper panel) and $\Theta$=0.785 (lower panel).}
\end{figure}
\begin{figure}
\centerline{\psfig{figure=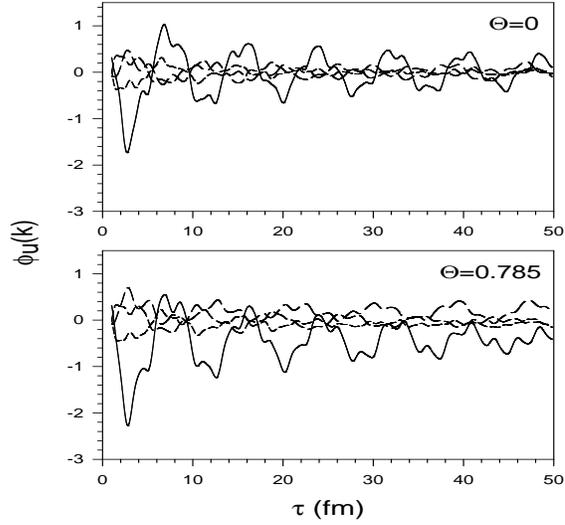,height=10cm,width=10cm}}
\caption{Evolution  of  the  Fourier transformed $\phi_u(k)$ with
time for different modes, k=6.2, 18.5, 30.9, 43.3 and  55.5  MeV.
The solid line is for the lowest mode.}
\end{figure}
\end{document}